\documentclass[aps,prl,twocolumn,groupedaddress]{revtex4}

\usepackage{graphics}
\usepackage{epsfig}
\renewcommand{\vec}[1]{\mbox{\boldmath $#1$}}

\begin{document}

\title{Critical Behaviour in the Relaminarisation of Localised
Turbulence in Pipe Flow}

\author{Ashley P. Willis}   
\email{A.Willis@bris.ac.uk}
\author{Rich R. Kerswell}
\email{R.R.Kerswell@bris.ac.uk}

\affiliation{Department of Mathematics, University of Bristol, University Walk,
   Bristol BS8 1TW, United Kingdom}

\date{\today}

\begin{abstract}

The statistics of the relaminarisation of localised turbulence in a
pipe are examined by direct numerical simulation.  As in recent
experimental data (Peixinho \& Mullin {\it Phys. Rev. Lett.} {\bf 96},
094501, 2006), the half life for the decaying turbulence is consistent
with the scaling $(Re_c-Re)^{-1}$, indicating a boundary crisis of the
localised turbulent state familiar in low-dimensional dynamical
systems.  The crisis Reynolds number, is estimated as $Re_c=1870$, a
value within $7\%$ of the experimental value $1750$. We argue that the
frequently-asked question of which $Re$ and initial disturbance are
needed to trigger sustained turbulence in a pipe, is really two
separate questions: the `local phase space' question (local to the
laminar state) of what threshold disturbance at a given $Re$ is needed
to initially trigger turbulence, followed by the `global phase space'
question of whether $Re$ exceeds $Re_c$ at which point the turbulent
state becomes an attractor.

\end{abstract}

\pacs{47.20.Ft,47.27.Cn,47.60.+i} 

\maketitle

\begin{figure*}
   \epsfig{figure=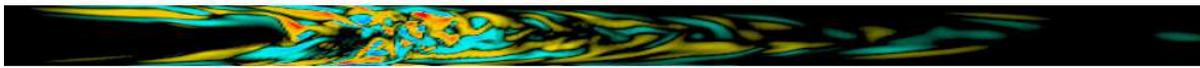, angle=0, width=160mm}
   \caption{\label{fig:puff}
      Numerical `puff' at $Re=1900$.  $(r,z)$-section of
      $(\vec{\nabla}\times\vec{u})_z$.  
      Only $20D$ shown of $50D$ computational domain.
   }
\end{figure*}

Understanding the behaviour of fluid flow through a circular straight
pipe remains one of the outstanding problems of classical physics and
has continued to intrigue the physics community for more than 160
years
\cite{hagen39},\cite{pois40},\cite{reynolds83},\cite{fitzgerald04}.
Although all evidence indicates that the laminar parabolic flow is
linearly stable, the flow can become turbulent even at modest flow
rates. The exact transition point depends not only on the flow rate
(measured by the Reynolds number $Re=UD/\nu$, where $U$ is the axial
flow speed, $D$ is the pipe diameter and $\nu$ is the fluid's
kinematic viscosity) but also sensitively on the shape and amplitude
of the disturbance(s) present \cite{darbyshire95}, \cite{hof03},
\cite{peixinho05}, \cite{peixinho06}. When it occurs, transition is
abrupt with the flow immediately becoming temporally and spatially
complex.  Given that most industrial pipe flows are turbulent and
hence more costly to power than if laminar, a central issue is
to understand the conditions which trigger sustained turbulence.
The problem is, however, severely complicated by the fact that the
threshold appears very sensitive to the exact form of the disturbance
{\it and} long turbulent transients can exist close to the threshold.
Of particular interest is the low-$Re$ situation where the transition
typically leads to a clearly localised turbulent structure called
a `puff' within the laminar flow \cite{reynolds83},
\cite{wygnanski73}. A puff has a typical length of about $20D$ along
the pipe (see Fig.\ \ref{fig:puff}) and, despite appearing
established, can relaminarise without warning after travelling many
hundreds of pipe diameters downstream.
%
%

There have been a number of contributions to this problem but so far
no consensus on the minimum Reynolds number, $Re_c$, above which
turbulence is sustained.  Experimental studies have focussed on
plotting transition-threshold curves in disturbance amplitude-$Re$
space for specific forms of applied perturbation.  One well-studied
perturbation having six-fold rotational symmetry gave rise to a
threshold amplitude which scaled like $Re^{-1}$ above $Re=2000$
\cite{hof03} but diverged at $Re_c \approx 1800$ \cite{peixinho05},
i.e. below this value no sustained turbulence could be excited however
hard the flow was disturbed.  Subsequent experiments \cite{peixinho06}
studying the statistics of relaminarisations of puffs as $Re$ is
reduced have lowered this threshold value to $Re_c=1750 \pm 10$, close
to a previous estimate of $1760$ \cite{darbyshire95} but not to others
of $1876$ \cite{gilbrech65} and $\approx 2000$ \cite{wygnanski73}.
The only complementary numerical work performed so far has been in a
short periodic pipe of $5D$ length \cite{faisst04} where it was
demonstrated that the pipe-long turbulent state displays the transient
characteristics of a chaotic repellor until $Re_c=2250$ above which it
becomes a chaotic attractor.  Recent experiments using a very long
pipe \cite{hof06} in which the statistics on long transients are
available, however, suggests that there is no critical
behaviour. Rather than the turbulent half life scaling like $\tau \sim
(Re_c-Re)^{-1}$ \cite{peixinho06},\cite{faisst04}, it is found to
increase exponentially instead. Interestingly, re-interpretation of
the $5D$-pipe data seems to corroborate this alternative exponential
lifetime behaviour even though the pipe is too short to capture a
turbulent puff.

In this Letter, we consider a much longer pipe of length $16 \pi D$ ($
\approx 50D$) in which turbulent puffs can be represented faithfully
using direct numerical simulation \cite{priymak04} and examine the
statistics of how they relaminarise. We find an exponential
distribution of lifetimes and the critical scaling law $\tau \sim
(Re_c-Re)^{-1}$, with a constant of proportionality and an estimate of
$Re_c=1870$ both in good agreement with experimental data
\cite{peixinho06}. Surprisingly, given its long history, this
represents the first time that a {\it quantitative} connection between
theory and experiment has been established in the pipe flow problem.


The Navier--Stokes equations for an incompressible
Newtonian fluid,
\begin{equation}
\partial_t \vec{u}+\vec{u}\cdot\vec{\nabla} \vec{u}+ \vec{\nabla}
 p=\nu\, \nabla^2 \vec{u}, \qquad \nabla\cdot\vec{u}=0,
\end{equation}
in a straight pipe with circular cross-section and for constant
mass-flux, were solved numerically in cylindrical coordinates
$(r,\theta,z)$ using a mixed pseudospectral-finite difference
formulation
\endnote{Incompressibility was satisfied automatically by adopting a
toroidal-poloidal potential formulation \cite{marques90}, further
reformulated into five simple second order equations in $r$.  The
numerical discretization was via a non-equispaced 9-point finite
difference stencil in $r$ and by Fourier modes in $\theta$ and $z$.
At the pipe wall boundary conditions coupling the potentials were
solved to numerical precision using an influence-matrix method, and
axial symmetry properties imposed by the geometry on each Fourier mode
were enforced implicitly in the finite difference weights.  }.
%
%
%
\begin{figure}
   \epsfig{figure=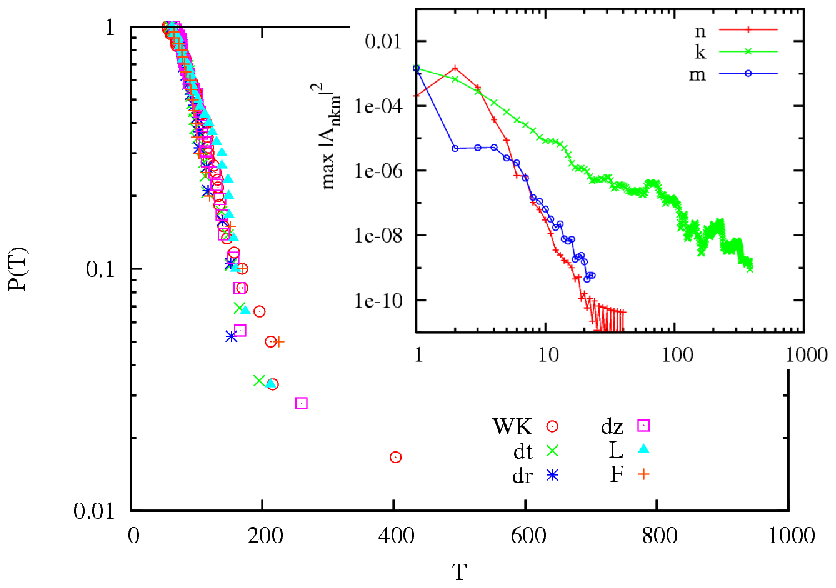, angle=0, width=85mm}
   \caption{ \label{fig:cgce2} Probability of relaminarisation after
   time T at $Re=1740$ is the same for increased resolutions (`dt' data with
   timestep halved, `dr' data with 60 radial points, `dz' data with
   axial resolution of $\pm 576$), pipe length (`L' is $100D$ data),
   and different disturbance (`F' is data obtained with the puff
   generated by an initial period of body forcing - the data is
   shifted in the last case to account for a longer transient
   period). All data sets effectively overlay the default data 'WK'.
   Inset, numerical puff spectrum at $Re=1900$,
   $A_n=\max_{km}|A_{nkm}|^2$, index $n$ of Chebyshev transformed
   radial modes, $k,m$ axial and azimuthal Fourier modes respectively
   and similarly for $A_k$ and $A_m$.}
\end{figure}
%
%
\begin{figure}
   \includegraphics{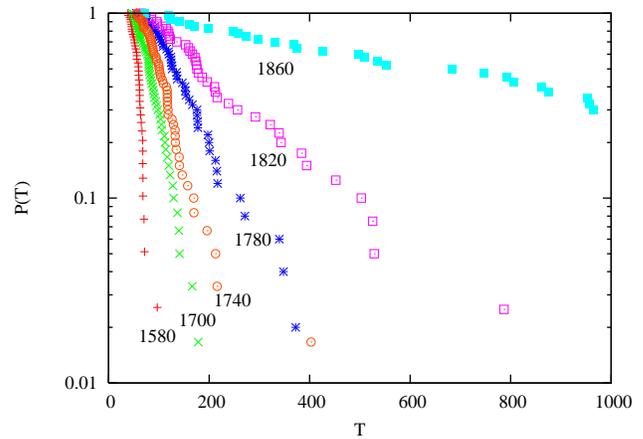}
   \caption{ \label{fig:PT} The probability of turbulent lifetime $
      \geq T$, $P(T)$, for several $Re$ in a periodic pipe of length
      $16 \pi D$.}
\end{figure}
The code was found to accurately reproduce linear stability results
for Hagen-Poiseuille flow, instabilities of nonlinear travelling wave
solutions and the statistical properties of turbulent pipe flow
\cite{eggels94} (as well as being cross-validated with another code
\cite{kerswell07}).  A resolution of 40 radial points was adopted with
grid points concentrated at the boundary, Fourier modes were kept up
to $\pm 24$ in $\theta$, and to $\pm 384$ in $z$ for a periodic pipe
of length $L=16\pi \,D$. This ensured spectral drop off of 6 orders of
magnitude in the power of the coefficients when representing a puff
velocity field at $Re=1900$: see inset in Fig.\ \ref{fig:cgce2}.  The
timestep was dynamically controlled using information from a
predictor-corrector method and was typically around $0.006\,D/U$. The
initial conditions for the calculations were randomly-selected
velocity snapshots taken from a long puff simulation performed at
$Re=1900$. A body forcing applied over $10D$ of the pipe and for a
time $10\,D/U$ was used to generate an `equilibrium' puff which
remained stable in length and form for a time period of over
$2000\,D/U$ (see Fig.\ \ref{fig:puff}). At a chosen $Re< 1900$ a
series of at least 40 and up to 60 independent simulations were
performed each initiated with a different puff snapshot to generate a
data set of relaminarisation times. The signature of the
relaminarisation was a clear and sudden transition to exponential
decay of the energy.  The criterion for relaminarisation was taken to
be such that the energy of the axially-dependent modes was less than
$5\times 10^{-4} \,\rho U^2 D^3$, below which all solutions were well
within the decaying regime.  
The range of measured $Re_c$ discussed above indicates sensitivity to noise.
Robustness of the relaminarisation
statistics was verified by comparing the half-lives of data sets
obtained by varying different computational parameters of the
simulation (see Fig.\ \ref{fig:cgce2}). All modifications produce
half-life values within the 95\% confidence interval about the
default half-life prediction.

\begin{figure}
   \epsfig{figure=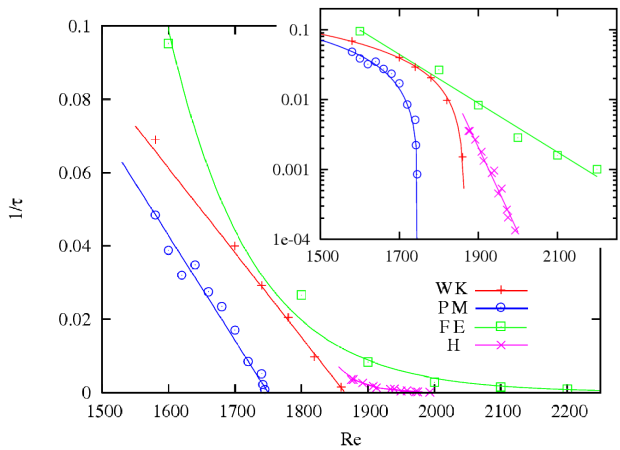, angle=0, width=85mm}
   \caption{ \label{fig:Retau1} The reciprocal of the puff half-life
   $\tau$ plotted against $Re$. Data plotted: `WK'- $50D$ data (each
   data point is the result of 40-60 simulations); `PM' - experimental
   data from \cite{peixinho06}; `FE' - reinterpreted $5D$ data
   \cite{faisst04}; `H' - experimental data from \cite{hof06}.  Inset,
   log-plot of $1/\tau$ vs $Re$. }
\end{figure}

Decay probabilities for a range of Reynolds numbers are shown in Fig.\
\ref{fig:PT} over an observation window of $1000 \,D/U$.  The linear
drop-off of the probability on the log-plot strongly suggests the
exponential distribution, $P(T) \sim \exp(-T\ln 2/\tau)$, where
$\tau=\tau(Re)$ is the half-life of a puff.  The median of $(T-t_0)$
was used as an estimator for $\tau$.  Inspection of the data by
varying the cut-off time $t_0$ revealed the effects of an initial
transient period in the first few data points. This was minimised by
selecting $t_0$ to exclude the first 5-10\% of the data (
determined by looking for the least sensitivity in the
half-life prediction).  The results plotted in Fig.\ \ref{fig:Retau1}
are consistent with the relation $\tau= \alpha(Re_c-Re)^{-1}$ where
$\alpha$ is $2.4 \times 10^{-4}$ compared to $2.8 \times 10^{-4}$
obtained in \cite{peixinho06} and there is a shift of 7\% in $Re_c$ up
to 1870 in the numerical data.  Also shown is the reinterpreted
numerical data for the $5D$ pipe \cite{faisst04} and the recent
half-life results from the long pipe experiments \cite{hof06} which
indicate that $1/\tau$ varies exponentially with $Re$ rather than
linearly.  Although the data from \cite{hof06} is for longer times,
there is sufficient overlap to suggest that the data from
\cite{peixinho06} and our results are not consistent with being the
earlier linear-looking part of this exponential. Rather, the results
indicate qualitatively different behaviour \footnote{In \cite{hof06},
the flow is disturbed by a jet of injected fluid much as in earlier
experiments \cite{peixinho05} where a six-jet disturbance was used.
This latter study found that results were sensitive to the exact flux
fraction of the laminar flow injected, with a (large) value of 0.1
giving $Re_c=1710 \pm 10$ whereas a (small) disturbance of 0.01 gave
$Re_c=1830 \pm 10$: \cite{hof06} quote injected flux rates of $\approx
0.07$. }.

%
%

The exponential probability distribution $P(T)$ found here in Fig.\
\ref{fig:PT} implies that puff relaminarisation is a {\em memoryless}
process - the probability that the puff will decay in a given interval
of time is proportional to the length of the period but independent of
previous events.  This feature has been found previously in turbulent
relaminarisation experiments in pipe flow
\cite{peixinho06},\cite{hof06},\cite{sreeni82} as well as in plane
Couette flow \cite{bottin98},\cite{bottinb98} and numerical
calculations using models of this together with other linearly stable
shear flows \cite{eckhardt02},\cite{moehlis04}.  Faisst and Eckhardt
\cite{faisst04} interpret this result as indicating that the transient
turbulent state for $Re<Re_c$ represents a chaotic repellor in phase
space.  Our results indicate that this conclusion carries over to a
localised turbulent puff in a long pipe. The building blocks for such
a repellor are saddle points and families of these in the form of
travelling waves with discrete rotational symmetries are now known to
exist down to $Re=1251$ \cite{faisst03}, \cite{wedin04},
\cite{kerswell05}. Tentative experimental evidence for their relevance
to puffs has already been found \cite{hof04} and corroborating
numerical evidence is now emerging \cite{kerswell07}.  The
entanglement of all the stable and unstable manifolds associated with
these saddles at some higher $Re$ presumably gives rise to
sufficiently complicated phase dynamics to appear as a turbulent puff
in real space. That this phase space structure is initially `leaky'
ultimately allowing escape (relaminarisation) is perhaps unsurprising
but what is less clear is how it suddenly becomes an attractor at
$Re_c$. The clean scaling of the transient decay half life, $\tau \sim
(Re_c-Re)^{-1}$ strongly suggests a boundary crisis \cite{grebogi86}
while the precise value of the critical exponent hints at a simple
dynamical systems explanation.  One, of course, cannot rule out the
possibility that the region never becomes an attractor with the exit
probability becoming extremely small but staying finite as $Re$
increases \cite{hof06}.  Or, in fact, that there are a number of
`leaks' which one by one seal up giving a half-life behaviour which
varies over a number of discrete time scales. Also, at some point, the
effect of noise must surely become significant 
over long
times.  However, the fact that the numerical simulations and the
experimental results \cite{peixinho06} are quantitatively consistent
despite being subject to different types of errors/disturbances
indicates that noise is not important over timescales of
$O(1000\,D/U)$ 
for the levels maintained here and in the
experiments.

\begin{figure}
   \epsfig{figure=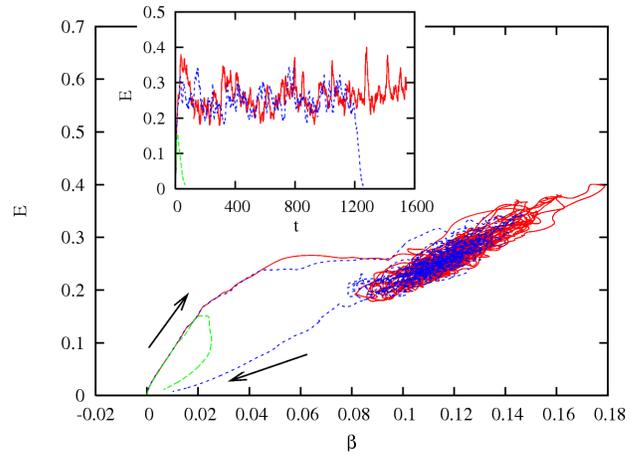, angle=0, width=85mm}
   \caption{ \label{fig:Ebeta} Trace of perturbation energy versus
   additional pressure fraction required to maintain fixed mass flux,
   $1+\beta = \, <\partial_z p> / \mathrm{d}_z p_{lam}$ (the origin
   represents laminar flow), for the three cases of a sustained puff
   at $Re=1900$ (solid), metastable puff at $Re=1860$ with sudden
   relaminarisation (dotted) and the immediate decay of a
   perturbation (dashed). The inset shows that the energy trace for
   the metastable puff is similar to the sustained puff before it
   laminarises.  }
\end{figure}
\begin{figure}
   \epsfig{figure=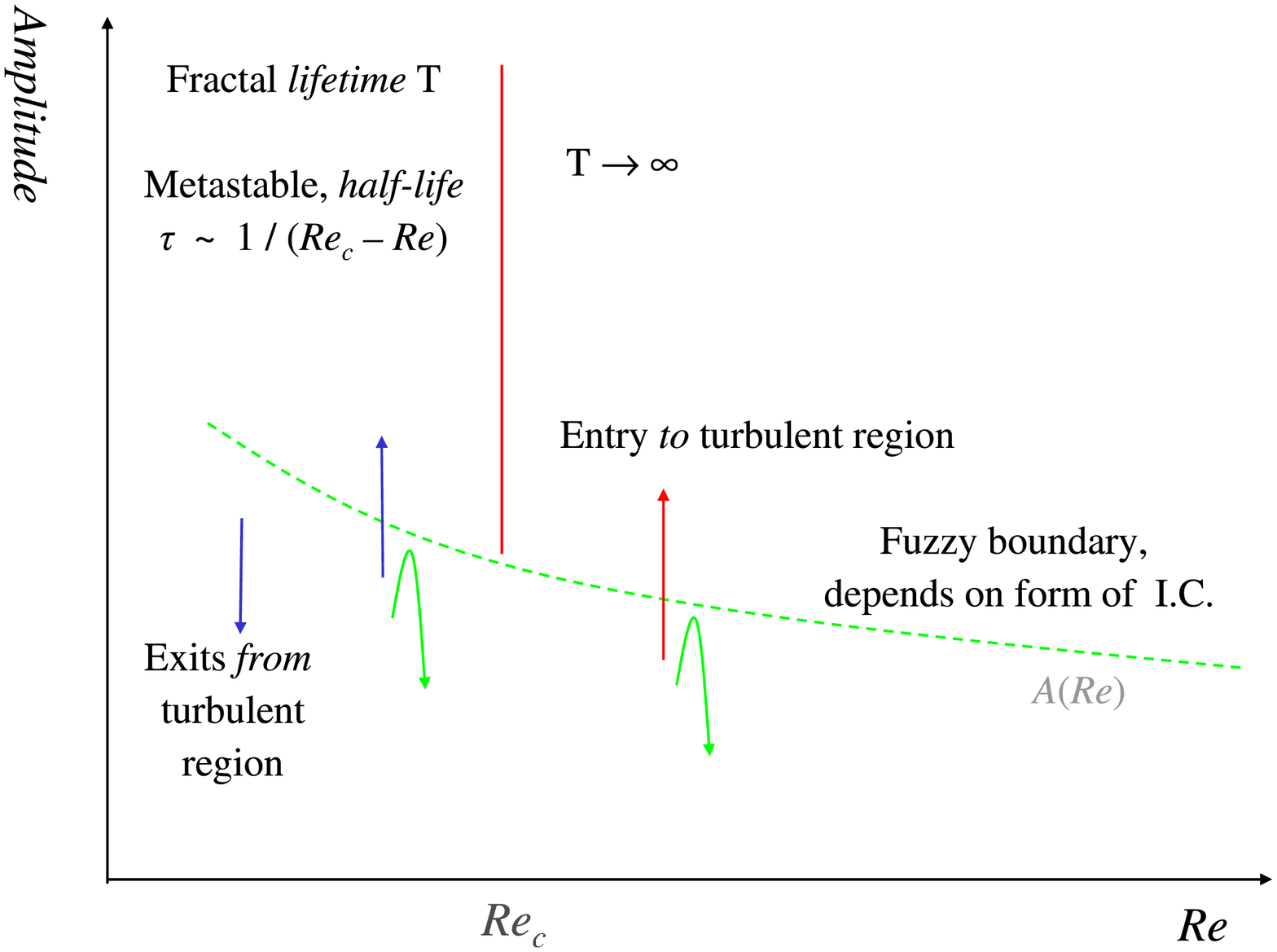, angle=270, width=80mm}
   \caption{\label{fig:thresh} Sketch of the two (independent)
   thresholds associated with transition: one is amplitude-dependent
   (and highly form-dependent) indicating when a turbulent episode is
   triggered, and the other is a global $Re$-dependent threshold
   indicating when the turbulence will be sustained.  }
\end{figure}

The simulations confirm that the puff characteristics are continuous
as $Re$ crosses $Re_c$ and that a puff corresponds to a part of
phase space disjoint from the laminar state (see Fig.\ \ref{fig:Ebeta}
and inset). This observation naturally divides the usual question as
to how to trigger sustained turbulence in pipe flow into two separate
issues.  Firstly, what disturbance at a given $Re$ is needed to
trigger turbulence initially --- i.e. what initial conditions will
cause the flow to leave the neighbourhood of the laminar state to
reach the puff region of phase space.  And secondly, what $Re$ is
needed so that, for a flow already in the turbulent region, the flow
never leaves --- i.e. the puff has become an attractor. The
implications of this realisation are that experimental curves in
\cite{darbyshire95} and \cite{peixinho05} showing a threshold curve on
a disturbance amplitude-$Re$ plot must, in fact, be two curves as
shown in Fig.\ \ref{fig:thresh}.
%
%
%
%
%
%
Figure \ref{fig:Ebeta} shows how initially a threshold amplitude of
disturbance is required to push the solution away from the laminar
state and into the turbulent region.  Once here, the exit from the
metastable state is sudden and unrelated to the entry as 
relaminarisation is a memoryless process.

%
%
To summarise, numerical simulations described in this Letter have
clarified the existence of two independent thresholds for sustained
turbulence.  Results probing the relaminarisation threshold closely
match a recent experimental investigation \cite{peixinho06}. For
timescales extending these experiments --- $t \leq 1000\,D/U$ --- we
confirm the presence of an exponential distribution for the
probability of puff relaminarisation and corroborate critical-type
behaviour in which the puff half-life diverges as $(Re_c-Re)^{-1}$.
Good quantitative agreement between the experimentally and
theoretically-estimated value of $Re_c$ (less than $7\%$ difference)
is a rare triumph in this famous canonical problem.

\begin{acknowledgments}
  We thank Jorge Peixinho, Tom Mullin and Bj\"{o}rn Hof for sharing
  their latest data prior to publication and, together with Bruno
  Eckhardt, for useful conversations. This research was funded by the
  EPSRC under grant GR/S76144/01.
\end{acknowledgments}


\begin{thebibliography}{26}
\expandafter\ifx\csname natexlab\endcsname\relax\def\natexlab#1{#1}\fi
\expandafter\ifx\csname bibnamefont\endcsname\relax
  \def\bibnamefont#1{#1}\fi
\expandafter\ifx\csname bibfnamefont\endcsname\relax
  \def\bibfnamefont#1{#1}\fi
\expandafter\ifx\csname citenamefont\endcsname\relax
  \def\citenamefont#1{#1}\fi
\expandafter\ifx\csname url\endcsname\relax
  \def\url#1{\texttt{#1}}\fi
\expandafter\ifx\csname urlprefix\endcsname\relax\def\urlprefix{URL }\fi
\providecommand{\bibinfo}[2]{#2}
\providecommand{\eprint}[2][]{\url{#2}}

\bibitem[{\citenamefont{Hagen}(1839)}]{hagen39}
\bibinfo{author}{\bibfnamefont{G.~H.~L.} \bibnamefont{Hagen}},
  \bibinfo{journal}{Poggendorfs Annalen der Physik und Chemie}
  \textbf{\bibinfo{volume}{16}}, \bibinfo{pages}{423} (\bibinfo{year}{1839}).

\bibitem[{\citenamefont{Poiseuille}(1840)}]{pois40}
\bibinfo{author}{\bibfnamefont{J.~L.~M.} \bibnamefont{Poiseuille}},
  \bibinfo{journal}{Comptes Rendus de l'Acad\'{e}mie des Sciences}
  \textbf{\bibinfo{volume}{11}}, \bibinfo{pages}{961,1041}
  (\bibinfo{year}{1840}).

\bibitem[{\citenamefont{Reynolds}(1883)}]{reynolds83}
\bibinfo{author}{\bibfnamefont{O.}~\bibnamefont{Reynolds}},
  \bibinfo{journal}{Proc. R. Soc. Lond.} \textbf{\bibinfo{volume}{35}},
  \bibinfo{pages}{84} (\bibinfo{year}{1883}).

\bibitem[{\citenamefont{Fitzgerald}(Feb 2004)}]{fitzgerald04}
\bibinfo{author}{\bibfnamefont{R.}~\bibnamefont{Fitzgerald}},
  \bibinfo{journal}{Physics Today}  (\bibinfo{year}{Feb 2004}).

\bibitem[{\citenamefont{Darbyshire and Mullin}(1995)}]{darbyshire95}
\bibinfo{author}{\bibfnamefont{A.~G.} \bibnamefont{Darbyshire}}
  \bibnamefont{and} \bibinfo{author}{\bibfnamefont{T.}~\bibnamefont{Mullin}},
  \bibinfo{journal}{J.\ Fluid Mech.} \textbf{\bibinfo{volume}{289}},
  \bibinfo{pages}{83} (\bibinfo{year}{1995}).

\bibitem[{\citenamefont{Hof et~al.}(2003)\citenamefont{Hof, Juel, and
  Mullin}}]{hof03}
\bibinfo{author}{\bibfnamefont{B.}~\bibnamefont{Hof}},
  \bibinfo{author}{\bibfnamefont{A.}~\bibnamefont{Juel}}, \bibnamefont{and}
  \bibinfo{author}{\bibfnamefont{T.}~\bibnamefont{Mullin}},
  \bibinfo{journal}{Phys.\ Rev.\ Lett.} \textbf{\bibinfo{volume}{91}},
  \bibinfo{pages}{244502} (\bibinfo{year}{2003}).

\bibitem[{\citenamefont{Peixinho and Mullin}(2005)}]{peixinho05}
\bibinfo{author}{\bibfnamefont{J.}~\bibnamefont{Peixinho}} \bibnamefont{and}
  \bibinfo{author}{\bibfnamefont{T.}~\bibnamefont{Mullin}},
  \bibinfo{journal}{Proc. IUTAM Symp. on Laminar-Turbulent Transition (eds
  Govindarajan, R. and Narasimha, R.)} pp. \bibinfo{pages}{45--55}
  (\bibinfo{year}{2005}).

\bibitem[{\citenamefont{Peixinho and Mullin}(2006)}]{peixinho06}
\bibinfo{author}{\bibfnamefont{J.}~\bibnamefont{Peixinho}} \bibnamefont{and}
  \bibinfo{author}{\bibfnamefont{T.}~\bibnamefont{Mullin}},
  \bibinfo{journal}{Phys.\ Rev.\ Lett.} \textbf{\bibinfo{volume}{96}},
  \bibinfo{pages}{094501} (\bibinfo{year}{2006}).

\bibitem[{\citenamefont{Wygnanski and Champagne}(1973)}]{wygnanski73}
\bibinfo{author}{\bibfnamefont{I.~J.} \bibnamefont{Wygnanski}}
  \bibnamefont{and} \bibinfo{author}{\bibfnamefont{F.~H.}
  \bibnamefont{Champagne}}, \bibinfo{journal}{J.\ Fluid Mech.}
  \textbf{\bibinfo{volume}{59}}, \bibinfo{pages}{281} (\bibinfo{year}{1973}).

\bibitem[{\citenamefont{Gilbrech and Hale}(1965)}]{gilbrech65}
\bibinfo{author}{\bibfnamefont{D.~A.} \bibnamefont{Gilbrech}} \bibnamefont{and}
  \bibinfo{author}{\bibfnamefont{J.~C.} \bibnamefont{Hale}},
  \emph{\bibinfo{title}{Further results on the transition from laminar to
  turbulent flow}} (\bibinfo{publisher}{Pergamon}, \bibinfo{year}{1965}),
  vol.~\bibinfo{volume}{2}, pp. \bibinfo{pages}{3--15}.

\bibitem[{\citenamefont{Faisst and Eckhardt}(2004)}]{faisst04}
\bibinfo{author}{\bibfnamefont{H.}~\bibnamefont{Faisst}} \bibnamefont{and}
  \bibinfo{author}{\bibfnamefont{B.}~\bibnamefont{Eckhardt}},
  \bibinfo{journal}{J.\ Fluid Mech.} \textbf{\bibinfo{volume}{504}},
  \bibinfo{pages}{343} (\bibinfo{year}{2004}).

\bibitem[{\citenamefont{Hof et~al.}(2006)\citenamefont{Hof, Westerweel,
  Schneider, and Eckhardt}}]{hof06}
\bibinfo{author}{\bibfnamefont{B.}~\bibnamefont{Hof}},
  \bibinfo{author}{\bibfnamefont{J.}~\bibnamefont{Westerweel}},
  \bibinfo{author}{\bibfnamefont{T.}~\bibnamefont{Schneider}},
  \bibnamefont{and} \bibinfo{author}{\bibfnamefont{B.}~\bibnamefont{Eckhardt}},
  \bibinfo{journal}{Nature in press} \textbf{\bibinfo{volume}{-}},
  (\bibinfo{year}{2006}).

\bibitem[{\citenamefont{Priymak and Miyazaki}(2004)}]{priymak04}
\bibinfo{author}{\bibfnamefont{V.~G.} \bibnamefont{Priymak}} \bibnamefont{and}
  \bibinfo{author}{\bibfnamefont{T.}~\bibnamefont{Miyazaki}},
  \bibinfo{journal}{Phys. Fluids} \textbf{\bibinfo{volume}{16}},
  \bibinfo{pages}{4221} (\bibinfo{year}{2004}).

\bibitem[{\citenamefont{Eggels et~al.}(1994)\citenamefont{Eggels, Unger, Weiss,
  Westerweel, Adrian, Friedrich, and Nieuwstadt}}]{eggels94}
\bibinfo{author}{\bibfnamefont{J.~G.~M.} \bibnamefont{Eggels}},
  \bibinfo{author}{\bibfnamefont{F.}~\bibnamefont{Unger}},
  \bibinfo{author}{\bibfnamefont{M.~H.} \bibnamefont{Weiss}},
  \bibinfo{author}{\bibfnamefont{J.}~\bibnamefont{Westerweel}},
  \bibinfo{author}{\bibfnamefont{R.~J.} \bibnamefont{Adrian}},
  \bibinfo{author}{\bibfnamefont{R.}~\bibnamefont{Friedrich}},
  \bibnamefont{and} \bibinfo{author}{\bibfnamefont{F.~T.~M.}
  \bibnamefont{Nieuwstadt}}, \bibinfo{journal}{J.\ Fluid Mech.}
  \textbf{\bibinfo{volume}{268}}, \bibinfo{pages}{175} (\bibinfo{year}{1994}).

\bibitem[{\citenamefont{Kerswell and Tutty}(2006)}]{kerswell07}
\bibinfo{author}{\bibfnamefont{R.~R.} \bibnamefont{Kerswell}} \bibnamefont{and}
  \bibinfo{author}{\bibfnamefont{O.}~\bibnamefont{Tutty}},
  \bibinfo{journal}{J.\ Fluid Mech. to be submitted}
  \textbf{\bibinfo{volume}{-}},  (\bibinfo{year}{2006}).

\bibitem[{\citenamefont{Sreenivasan}(1982)}]{sreeni82}
\bibinfo{author}{\bibfnamefont{K.~R.} \bibnamefont{Sreenivasan}},
  \bibinfo{journal}{Acta Mechanica} \textbf{\bibinfo{volume}{44}},
  \bibinfo{pages}{1} (\bibinfo{year}{1982}).

\bibitem[{\citenamefont{Bottin and Chate}(1998)}]{bottin98}
\bibinfo{author}{\bibfnamefont{S.}~\bibnamefont{Bottin}} \bibnamefont{and}
  \bibinfo{author}{\bibfnamefont{H.}~\bibnamefont{Chate}},
  \bibinfo{journal}{Eur. Phys. J. B} \textbf{\bibinfo{volume}{6}},
  \bibinfo{pages}{143} (\bibinfo{year}{1998}).

\bibitem[{\citenamefont{Bottin et~al.}(1998)\citenamefont{Bottin, Daviaud,
  Manneville, and Dauchot}}]{bottinb98}
\bibinfo{author}{\bibfnamefont{S.}~\bibnamefont{Bottin}},
  \bibinfo{author}{\bibfnamefont{F.}~\bibnamefont{Daviaud}},
  \bibinfo{author}{\bibfnamefont{P.}~\bibnamefont{Manneville}},
  \bibnamefont{and} \bibinfo{author}{\bibfnamefont{O.}~\bibnamefont{Dauchot}},
  \bibinfo{journal}{Europhys. Lett.} \textbf{\bibinfo{volume}{43}},
  \bibinfo{pages}{171} (\bibinfo{year}{1998}).

\bibitem[{\citenamefont{Eckhardt et~al.}(2002)\citenamefont{Eckhardt, Faisst,
  Schmiegel, and Schumacher}}]{eckhardt02}
\bibinfo{author}{\bibfnamefont{B.}~\bibnamefont{Eckhardt}},
  \bibinfo{author}{\bibfnamefont{H.}~\bibnamefont{Faisst}},
  \bibinfo{author}{\bibfnamefont{A.}~\bibnamefont{Schmiegel}},
  \bibnamefont{and}
  \bibinfo{author}{\bibfnamefont{J.}~\bibnamefont{Schumacher}},
  \bibinfo{journal}{Advances in Turbulence IX: Proceedings of the Ninth
  European Turbulence Conference, Barcelona edited by I.P.Castro, P.E. Hancock
  and T.G.Thomas} p. \bibinfo{pages}{701} (\bibinfo{year}{2002}).

\bibitem[{\citenamefont{Moehlis et~al.}(2004)\citenamefont{Moehlis, Faisst, and
  Eckhardt}}]{moehlis04}
\bibinfo{author}{\bibfnamefont{J.}~\bibnamefont{Moehlis}},
  \bibinfo{author}{\bibfnamefont{H.}~\bibnamefont{Faisst}}, \bibnamefont{and}
  \bibinfo{author}{\bibfnamefont{B.}~\bibnamefont{Eckhardt}},
  \bibinfo{journal}{New Journal of Physics} \textbf{\bibinfo{volume}{6}},
  \bibinfo{pages}{56} (\bibinfo{year}{2004}).

\bibitem[{\citenamefont{Faisst and Eckhardt}(2003)}]{faisst03}
\bibinfo{author}{\bibfnamefont{H.}~\bibnamefont{Faisst}} \bibnamefont{and}
  \bibinfo{author}{\bibfnamefont{B.}~\bibnamefont{Eckhardt}},
  \bibinfo{journal}{Phys. Rev. Lett.} \textbf{\bibinfo{volume}{91}},
  \bibinfo{pages}{224502} (\bibinfo{year}{2003}).

\bibitem[{\citenamefont{Wedin and Kerswell}(2004)}]{wedin04}
\bibinfo{author}{\bibfnamefont{H.}~\bibnamefont{Wedin}} \bibnamefont{and}
  \bibinfo{author}{\bibfnamefont{R.~R.} \bibnamefont{Kerswell}},
  \bibinfo{journal}{J.\ Fluid Mech.} \textbf{\bibinfo{volume}{508}},
  \bibinfo{pages}{333} (\bibinfo{year}{2004}).

\bibitem[{\citenamefont{Kerswell}(2005)}]{kerswell05}
\bibinfo{author}{\bibfnamefont{R.~R.} \bibnamefont{Kerswell}},
  \bibinfo{journal}{Nonlinearity} \textbf{\bibinfo{volume}{18}},
  \bibinfo{pages}{R17} (\bibinfo{year}{2005}).

\bibitem[{\citenamefont{Hof and et~al}(2004)}]{hof04}
\bibinfo{author}{\bibfnamefont{B.}~\bibnamefont{Hof}} \bibnamefont{and}
  \bibinfo{author}{\bibnamefont{et~al}}, \bibinfo{journal}{Science}
  \textbf{\bibinfo{volume}{305}}, \bibinfo{pages}{1594} (\bibinfo{year}{2004}).

\bibitem[{\citenamefont{Grebogi et~al.}(1986)\citenamefont{Grebogi, Ott, and
  Yorke}}]{grebogi86}
\bibinfo{author}{\bibfnamefont{C.}~\bibnamefont{Grebogi}},
  \bibinfo{author}{\bibfnamefont{E.}~\bibnamefont{Ott}}, \bibnamefont{and}
  \bibinfo{author}{\bibfnamefont{J.}~\bibnamefont{Yorke}},
  \bibinfo{journal}{Phys. Rev. Lett.} \textbf{\bibinfo{volume}{57}},
  \bibinfo{pages}{1284} (\bibinfo{year}{1986}).

\bibitem[{\citenamefont{Marqu\'es}(1990)}]{marques90}
\bibinfo{author}{\bibfnamefont{F.}~\bibnamefont{Marqu\'es}},
  \bibinfo{journal}{Phys.\ Fluids A} \textbf{\bibinfo{volume}{2}},
  \bibinfo{pages}{729} (\bibinfo{year}{1990}).

\end{thebibliography}

\end{document}